\documentstyle[aps,preprint,epsf]{revtex}

\begin{document}
\draft \tightenlines

\preprint{}

\title{Quantum Field Theory in a Topology Changing Universe}

\author{Sang Pyo Kim\footnote{Electronic address:
sangkim@knusun1.kunsan.ac.kr}}

\address{Department of Physics,
Kunsan National University,
Kunsan 573-701, Korea}
\date{\today}

\maketitle
\begin{abstract}
We propose a method to construct quantum theory of matter fields
in a topology changing universe. Analytic continuation of the
semiclassical gravity of a Lorentzian geometry leads to a
non-unitary Schr\"{o}dinger equation in a Euclidean region of
spacetime, which does not have a direct interpretation of quantum
theory of the Minkowski spacetime. In this Euclidean region we
quantize the Euclidean geometry, derive the time-dependent
Schr\"{o}dinger equation and find the quantum states using the
Liouville-Neumann method. The Wick rotation of these quantum
states provides the correct Hilbert space of matter field in the
Euclidean region of the Lorentzian geometry. It is found that the
direct quantization of a scalar field in the Lorentzian geometry
involves an unusual commutation rule in the Euclidean region.
Finally we discuss the interpretation of the periodic solution of
the semiclassical gravity equation in the Euclidean geometry as a
finite temperature solution for the gravity-matter system in the
Lorentzian geometry.
\end{abstract}

\section{Introduction}

The open inflation model proposed recently by Hawking-Turok
\cite{hawking-turok} revived an interest in the wave functions of
quantum cosmology, and provoked a debate on the boundary
conditions of the Universe \cite{linde-hawking}. The present
Lorentzian spacetime is supposed to emerge quantum mechanically
from a Euclidean spacetime. Leading proposals for such wave
functions are the Hartle-Hawking's no-boundary wave function
\cite{hartle-hawking}, the Linde's wave function \cite{linde} and
the Vilenkin's tunneling wave function \cite{vilenkin}. Though
there has not still been an unanimous agreement on the boundary
condition of the Universe, it is generally accepted that the
Universe which tunneled quantum mechanically the Euclidean
spacetime should have either an exponentially growing or decaying
wave function or a combination of both branches \cite{vilenkin2}.
But the issue of quantum theory of matter fields in a topology
changing universe such as the tunneling universe has not been
raised seriously yet.

It is the purpose of this paper to investigate a consistent
quantum theory of a scalar field in the Euclidean region of the
topology changing universe such as the tunneling universe and to
construct quantum states explicitly there. To treat the quantum
field in a curved spacetime there have been used two typical
methods: in the conventional approach the underlying spacetime is
fixed as a background and matter field is quantized on it
\cite{birrel}, and in the other approach known as the
semiclassical (quantum) gravity the semiclassical Einstein
equation with the quantum back-reaction of matter field and the
time-dependent Schr\"{o}dinger equation are derived from the
Wheeler-DeWitt equation for the gravity-matter system
\cite{kiefer,brout,backreact,kim}. In both approaches the
functional Schr\"{o}dinger equation for the scalar field in the
Euclidean region obeys a non-unitary (diffusion-like) evolution
equation. So the quantization rules of the Minkowski spacetime may
not be applied directly. To construct the consistent quantum
theory in the Euclidean region we propose a method in which one
first quantizes the Wick-rotated gravity-matter system of the
Euclidean geometry, derives the time-dependent Schr\"{o}dinger
equation and then transforms back via the Wick rotation the
quantum states into those of the Lorentzian geometry. This method
is consistent because the Wick rotations are well defined and so
does the time-dependent Schr\"{o}dinger equation in the Euclidean
region in the Euclidean geometry just as in the Lorentzian region
of the Lorentzian geometry. It is also useful in that one is able
to find the quantum states explicitly using the Liouville-Neumann
method which has already been used to find quantum states of the
scalar field in the Lorentzian regions of spacetime \cite{kim2}.

The organization of the paper is as follows. In Sec. II we
quantize the gravity-scalar system in the Lorentzian geometry and
derive the semiclassical Einstein equation and the time-dependent
Schr\"{o}dinger equation from the Wheeler-DeWitt equation in the
Lorentzian region where the gravitational wave function
oscillates. The region where the gravitational wave function
exhibits an exponential behavior corresponds to a Euclidean region
of spacetime. We focus in particular on the Schr\"{o}dinger
equation in the Euclidean region of spacetime. In Sec. III we
quantize the Euclidean gravity (geometry) coupled to the minimal
scalar field and derive the time-dependent Schr\"{o}dinger
equation together with the semiclassical Einstein equation in the
region corresponding to the Euclidean region of the Lorentzian
geometry. Quantum states are found using the Liouville-Neumann
method. In Sec. IV the Wick rotation is employed to transform
these quantum states defined in terms of the Euclidean geometry
into those defined in terms of the Lorentzian geometry.

\section{Quantum Theory in Lorentzian Geometry}

As a simple but interesting quantum cosmological model, let us
consider the closed FRW universe minimally coupled to an inflaton,
a minimal scalar field. The action for the gravity with a
cosmological constant $\Lambda$ and the scalar field takes the
form
\begin{equation}
I = \frac{m_P^2}{16 \pi} \int d^4x \sqrt{-g} \Bigl[ R - 2 \Lambda
\Bigr]  + \frac{m_P^2}{8 \pi} \int d^3x \sqrt{h} K + \int d^4x
\sqrt{-g} \Bigl[\frac{1}{2} g^{\mu \nu}
\partial_{\mu} \phi \partial_{\nu} \phi - V(\phi)  \Bigr],
\end{equation}
where $m_P^2 = 1/G$ is the Planck mass squared. The surface term
for the gravity has been introduced to yield the correct Einstein
equation for the closed universe. In the Lorentzian FRW universe
with the metric in the ADM formulation
\begin{equation}
ds^2_L = - N^2(t) dt^2 + a^2 (t) d \Omega_3^2, \label{metric1}
\end{equation}
the action becomes
\begin{equation}
I_L = \int dt \Biggl[ - \frac{3 \pi m_P^2}{4} \Biggl( \frac{a}{N}
\Bigl( \frac{\partial a}{\partial t}\Bigr)^2
 - N V_g (a)  \Biggr)
+  2 \pi^2 a^3 \Biggl( \frac{1}{2 N} \Bigl(\frac{\partial \phi}{\partial t}
\Bigr)^2 - N V (\phi) \Biggr)
\Biggr],
\label{lorentz act}
\end{equation}
where
\begin{equation}
V_g (a) = a - \frac{\Lambda}{3} a^3.
\end{equation}
In the above equation we dropped the second order derivative term,
which is to be cancelled by a boundary action. By introducing the
canonical momenta
\begin{equation}
\pi_a = - \frac{3 \pi m_P^2 a}{2 \pi N} \frac{\partial a}{\partial
t}, ~ \pi_{\phi} = \frac{2 \pi^2 a^3}{N} \frac{\partial
\phi}{\partial t},
\end{equation}
one obtains the Hamiltonian constraint
\begin{equation}
{\cal H}_L = \hat{H}_g (\pi_a, a) + \hat{H}_L (\pi_{\phi}, \phi,
a) = 0,
\end{equation}
where
\begin{eqnarray}
\hat{H}_g (\pi_a, a) &=& - \frac{1}{3 \pi m_P^2 a} \pi_a^2 -
\frac{3 \pi m_P^2}{4} V_g (a), \\ \hat{H}_L (\pi_{\phi}, \phi, a)
&=& \frac{1}{4 \pi^2 a^3} \pi_{\phi}^2 + 2 \pi^2 a^3 V(\phi)
\end{eqnarray}
are the Hamiltonians for the gravity and scalar field,
respectively. The Dirac quantization leads to the Wheeler-DeWitt
equation for the Lorentzian geometry
\begin{equation}
\Biggl[ - \frac{\hbar^2}{3 \pi m_P^2 a} \frac{\partial^2}{\partial
a^2} + \frac{3 \pi m_P^2}{4} V_g (a) + \frac{\hbar^2}{4 \pi^2 a^3}
\frac{\partial^2}{\partial {\phi}^2} - 2 \pi^2 a^3 V(\phi) \Biggr]
\Psi_L(a, \phi) = 0, \label{lwd eq}
\end{equation}
where we neglected the operator ordering ambiguity.

Before we derive the equation for quantum fields in the Euclidean
region, we review briefly how to obtain the time-dependent
Schr\"{o}dinger equation in the Lorentzian region from the
semiclassical (quantum) gravity point of view. In the Lorentzian
region, where the wave function for the gravitational field
oscillates, we first adopt the Born-Oppenheimer idea to expand the
total wave function according to different mass scales
\begin{equation}
\Psi_L (a, \phi) = \psi_L (a) \Phi_L (\phi, a), \label{tot wav}
\end{equation}
and obtain the gravitational field equation with the back-reaction
of matter field \cite{backreact}
\begin{equation}
\Biggl[ - \frac{\hbar^2}{3 \pi m_P^2 a} D^2 + \frac{3 \pi
m_P^2}{4} V_g (a) + \langle \hat{H}_L \rangle - \frac{\hbar^2}{3
\pi m_P^2 a} \langle \bar{D}^2 \rangle \Biggr] \psi_L(a) = 0.
\label{grav eq}
\end{equation}
Here, $D$ and $\bar{D}$ denote the covariant derivatives
\begin{equation}
D = \frac{\partial}{\partial a} + i A (a), ~~ \bar{D} =
\frac{\partial}{\partial a} - i A (a),
\end{equation}
with an effective gauge potential from the scalar field
\begin{equation}
A (a) = - i \frac{\langle \Phi_L \vert \frac{\partial}{\partial a
} \vert \Phi_L \rangle}{\langle \Phi_L \vert \Phi_L \rangle},
\end{equation}
and $\langle \hat{H}_L \rangle$ and $\langle \bar{D}^2 \rangle$
denote the expectation value of the corresponding operators
\begin{equation}
\langle \hat{H}_L \rangle = \frac{\langle \Phi_L \vert \hat{H}_L
\vert \Phi_L \rangle}{\langle \Phi_L \vert \Phi_L \rangle},~~
\langle \bar{D}^2 \rangle = \frac{\langle \Phi_L \vert \hat{H}_L
\vert \Phi_L \rangle}{\langle \Phi_L \vert \Phi_L \rangle}.
\end{equation}
By putting Eq. (\ref{tot wav}) into Eq. (\ref{lwd eq}) and by
subtracting Eq. (\ref{grav eq}), one gets the equation for the
matter field
\begin{equation}
- \frac{2 \hbar^2}{3 \pi m_P^2 a} \frac{1}{\psi_L} \bigl(D \psi_L
\bigr) \bigl(\bar{D} \Phi_L \bigr) + \bigl( \hat{H}_L - \langle
\hat{H}_L \rangle \bigr) \Phi_L -  \frac{\hbar^2}{3 \pi m_P^2 a}
\bigl(\bar{D}^2 - \langle \bar{D}^2 \rangle \bigr) \Phi_L = 0.
\label{mat eq}
\end{equation}
Since $A(a)$ is a gauge potential and a $c$-number, the geometric
phases for the wave function and quantum state
\begin{equation}
\psi_L (a) = e^{ - i \int A} \tilde{\psi}_L, ~~ \Phi_L = e^{i \int
A} \tilde{\Phi}_L, \label{gauge wav}
\end{equation}
remove the gauge potentials from the covariant derivatives $D$ and
$\bar{D}$, in Eqs. (\ref{grav eq}) and (\ref{mat eq}). However,
the total wave function (\ref{tot wav}) keeps the same form
$\Psi_L = \tilde{\psi}_L \tilde{\Phi}_L$. From now on we shall
work with the wave function and quantum state (\ref{gauge wav}),
drop the tildes for simplicity and ignore the last terms in Eqs.
(\ref{grav eq}) and (\ref{mat eq}), which are small compared with
the other terms.

We then follow the de Broglie-Bohm interpretation and set the
gravitational wave function in the form
\begin{equation}
\psi_{L(II)} (a) = F(a) \exp \Bigl[\pm \frac{i}{\hbar} S_{L(II)}
(a) \Bigr], \label{db1}
\end{equation}
Here, $(II)$ denotes an oscillatory region of Lorentzian geometry
(see Fig.1) and $\pm$ signs correspond to the expanding and
collapsing branches of the universe, respectively. The real part
gives rise to the Hamilton-Jacobi equation
\begin{equation}
  \frac{1}{3 \pi m_P^2 a} \Bigl( \frac{\partial S_{L(II)}}{\partial a}
 \Bigr)^2 + \frac{3 \pi m_P^2}{4} V_g  - \frac{1}{3 \pi m_P^2 a}
 V_q - \langle \hat{H}_m \rangle  = 0,
 \label{sem ein}
\end{equation}
where
\begin{equation} V_q (a) = \hbar^2 \frac{\partial^2
F/\partial a^2}{F}
\end{equation}
is the quantum potential. The imaginary part leads to the continuity equation
\begin{equation}
F \frac{\partial^2 S_{L(II)}}{\partial a^2} + 2 \frac{\partial F}{\partial a}
\frac{\partial S_{L(II)}}{\partial a} = 0.
\end{equation}
The contribution $V_q$ from the quantum potential will also be
ignored, which is at most one-loop or higher orders. By
integrating $c$-number $\langle \hat{H}_L \rangle$ and writing it
as a phase factor of $\Phi_L = e^{i \int \langle \hat{H}_L
\rangle} \tilde{\tilde{\Phi}}_L$ in Eq. (\ref{mat eq}) and once
again dropping the tilde for simplicity, one also obtains the
time-dependent Schr\"{o}dinger equation for the scalar field
\begin{equation}
i \hbar \frac{\partial}{\partial t} \Phi_L (\phi, t) = \hat{H}_L
\Bigl(\frac{\hbar}{i} \frac{\partial}{\partial \phi}, \phi, t
\Bigr) \Phi_L (\phi, t),
\end{equation}
where $t$ is the cosmological (WKB) time
\begin{equation}
\frac{\partial}{\partial t} = \mp \frac{2}{3 \pi m_P^2 a}
\frac{\partial S_{L(II)}}{\partial a} \frac{\partial}{\partial a}.
\label{cos time1}
\end{equation}
By identifying the cosmological time (\ref{cos time1}) with the
comoving time in Eq. (\ref{metric1}) and by making use of
\begin{equation}
\frac{\partial S_{L(II)}}{\partial a} = \mp \frac{3 \pi m_P^2
a}{2} \frac{\partial a}{\partial t},
\end{equation}
one sees that Eq. (\ref{sem ein}) becomes indeed the semiclassical
Einstein equation
\begin{equation}
\Biggl(\frac{\frac{\partial a}{\partial t}}{a} \Biggr)^2 +
\frac{1}{a^2} - \frac{\Lambda}{3} = \frac{4}{3 \pi m_P^2 a^3}
\langle \hat{H}_{L} \rangle. \label{sem ein1-1}
\end{equation}
The spacetime regions are divided according as the effective
potential for the gravitational field
\begin{equation}
V_{L} (a) =  \frac{3 \pi m_P^2}{4} V_g (a) - \langle \hat{H}_{L}
\rangle \label{leff pot}
\end{equation}
takes positive or negative values. For the sake of simplicity, we
assume that the quantum back-reaction of the scalar field is
insignificant compared with $V_g$, so that the effective potential
$V_L$ has a simple form in Fig. 1. The region I of Fig. 1, where
$V_{L}$ is positive, corresponds to a part of Euclidean spacetime,
whereas the region II, where $V_{L}$ is negative, corresponds to a
part of Lorentzian spacetime.

Being mostly interested in the quantum creation of the universe
from the Euclidean region of the tunneling universe to the
Lorentzian region, we focus on the region I of Fig. 1. Though the
gravitational motion is prohibited classically in the region I, it
is, however, permitted quantum mechanically. In this region one is
tempted to continue analytically the wave function (\ref{db1}) to
get
\begin{equation}
\psi_{L(I)} (a) = F(a) \exp \Bigl[\mp \frac{1}{\hbar} S_{L(I)} (a)
\Bigr], \label{grav funI}
\end{equation}
whose dominant contribution to Eq. (\ref{grav eq}) leads to the
Hamilton-Jacobi-like equation
\begin{equation}
\Bigl( \frac{\partial S_{L(I)}}{\partial a}
 \Bigr)^2 =   3\pi m_P^2 a V_{L}.
 \label{sem einI}
\end{equation}
At the same time one is able to obtain from Eq. (\ref{mat eq}) the
time-dependent Schr\"{o}dinger equation
\begin{equation}
\hbar \frac{\partial}{\partial s} \Phi_L (\phi, s)
= \hat{H}_{L} \Bigl(\frac{\hbar}{i}
\frac{\partial}{\partial \phi}, \phi, s \Bigr)
\Phi_L (\phi, s),
\label{sch eqI}
\end{equation}
where $s$ is a Euclidean analog of cosmological time defined by
\begin{equation}
\frac{\partial}{\partial s} = \pm \frac{2}{3 \pi m_P^2 a}
\frac{\partial S_{L(I)}}{\partial a} \frac{\partial}{\partial a}.
\end{equation}
But the scalar field Hamiltonian $\hat{H}_{L}$ keeps the same
form.

Then the following questions are raised. What is the meaning of
the parameter $s$? Is it an analytic continuation of the
cosmological time or a Wick-rotated Euclidean time? How to solve
Eq. (\ref{sch eqI}), an apparently time-dependent diffusion-like
equation? What are the quantization rule $\bigl[\hat{\phi},
\hat{\pi}_{\phi} \bigr]$ and the meaning of quantum states of this
non-unitary evolution? To answer these questions and to follow
analogy with quantum theory of Lorentzian spacetime, we shall
consider the quantum theory of the scalar field by quantizing the
Euclidean geometry.

\section{Quantum Theory in Euclidean Geometry}

To obtain the quantum cosmological model for the Euclidean
spacetime, we perform the Wick rotation
$t =  i \tau $ and consider the Euclidean metric
\begin{equation}
ds^2 = N^2 (\tau) d\tau^2 + a^2 (\tau) d \Omega^2_3.
\label{metric2}
\end{equation}
From the Euclidean action
\begin{equation}
I_E =  \int d\tau \Biggl[  \frac{3 \pi m_P^2}{4} \Biggl(
\frac{a}{N} \Bigl(\frac{\partial a}{\partial \tau} \Bigr)^2
 + N V_g (a) \Biggr)
- 2 \pi^2 a^3 (\tau) \Biggl( \frac{1}{2 N} \Bigl(
\frac{\partial \phi}{\partial \tau} \Bigr)^2 + N V (\phi) \Biggr)
\Biggr],
\label{eact}
\end{equation}
we obtain the Hamiltonian constraint
\begin{equation}
{\cal H}_E =  \frac{1}{3 \pi m_P^2 a} \pi_{E,a}^2 - \frac{3 \pi
m_P^2}{4} V_g (a) - \frac{1}{4 \pi^2 a^3} \pi_{E, \phi}^2 + 2
\pi^2 a^3 V(\phi) = 0, \label{econst}
\end{equation}
where
\begin{equation}
\pi_{E,a} =  \frac{3 \pi m_P^2 a}{2 N} \frac{\partial a}{\partial
\tau}, ~ \pi_{E, \phi} = - \frac{2 \pi^2 a^3}{N} \frac{\partial
\phi}{\partial \tau}.
\end{equation}
The Hamiltonian constraint (\ref{econst}) leads to the
Wheeler-DeWitt equation for the Euclidean geometry
\begin{equation}
\Biggl[  - \frac{\hbar^2}{3 \pi m_P^2 a}
\frac{\partial^2}{\partial a^2} - \frac{3 \pi m_P^2}{4} V_g (a) +
\frac{\hbar^2}{4 \pi^2 a^3} \frac{\partial^2}{\partial {\phi}^2} +
2 \pi^2 a^3 V(\phi) \Biggr] \Psi_E(a, \phi) = 0. \label{ewd eq}
\end{equation}
It should be remarked that the Wick rotation changed both signs of
the kinetic terms of the scalar and gravitational fields. This is
the reason why the Euclidean action can not be made positive
definite for the gravity-matter system.

We now wish to obtain the semiclassical Einstein equation and the
time-dependent Schr\"{o}dinger equation in the Euclidean region I
of Fig. 1. Since the sign of the kinetic term of gravitational
field was reversed due to the Wick rotation, the wave function
$\Psi_E$ of the Wheeler-DeWitt equation (\ref{ewd eq}) now
oscillates in the region I. This is in contrast with the behavior
of the wave function $\Psi_L$. Thus we may use the semiclassical
quantum gravity approach in Sec. II and Ref. \cite{kim}.  As in
the Lorentzian spacetime we may expand the total wave function in
the form of Eq. (\ref{tot wav}) and set the wave function for the
gravity in the form
\begin{equation}
\psi_E (a) = F(a) \exp \Bigl[\pm \frac{i}{\hbar} S_E (a) \Bigr].
\end{equation}
The real part of the resultant gravitational field equation, which
is similar to Eq. (\ref{grav eq}) with the reversed signs for both
kinetic terms, gives rise to the Hamilton-Jacobi equation
\begin{equation}
 \frac{1}{3 \pi m_P^2 a} \Bigl( \frac{\partial S_E}{\partial a}
 \Bigr)^2 - \frac{3 \pi m_P^2}{4} V_g + \langle \hat{H}_{E} \rangle  = 0,
 \label{sem ein2}
\end{equation}
where
\begin{equation}
\hat{H}_{E} =  \frac{\hbar^2}{4 \pi^2 a^3} \frac{\partial^2}{\partial {\phi}^2}
+ 2 \pi^2 a^3 V(\phi)
\end{equation}
is the scalar field Hamiltonian in the Euclidean geometry. As in
the Lorentzian geometry, one can get the semiclassical Einstein
equation in the Euclidean geometry
\begin{equation}
\Biggl(\frac{\frac{\partial a}{\partial \tau}}{a} \Biggr)^2 -
\frac{1}{a^2} + \frac{\Lambda}{3} = - \frac{4}{3 \pi m_P^2 a^3}
\langle \hat{H}_{E} \rangle. \label{sem ein2-2}
\end{equation}

We note that the Euclidean region I corresponds to the region
where the effective potential for the gravitational field
\begin{equation}
V_{E} (a) =  \frac{3 \pi m_P^2}{4} V_g - \langle \hat{H}_{E}
\rangle
\end{equation}
takes positive values. To consolidate this point further, let us
remind that $\langle \hat{H}_{E} \rangle$ is a Wick rotation of
$\langle \hat{H}_{L} \rangle$, as will be shown later. So the
region where $V_{E}$ is positive, coincides with the region I
where $V_{L}$ is positive, too. However, the wave function for the
quantized Euclidean geometry oscillates in this region, in
contrast with the exponential behavior of the wave function for
the quantized Lorentzian geometry. Therefore, we are able to
obtain the time-dependent unitary Schr\"{o}dinger equation
\begin{equation}
i \hbar \frac{\partial}{\partial \tau}
\Phi_E (\phi, \tau) = \hat{H}_{E} \Bigl(\frac{\hbar}{i}
\frac{\partial}{\partial \phi}, \phi, \tau \Bigr)
\Phi_E (\phi, \tau),
\label{esch eq}
\end{equation}
where $\tau$ is the cosmological time defined as
\begin{equation}
\frac{\partial}{\partial \tau} = \mp \frac{2}{3 \pi m_P^2 a}
\frac{\partial S_E}{\partial a} \frac{\partial}{\partial a}.
\label{cos time2}
\end{equation}
The cosmological time (\ref{cos time2}) coincides with the
Euclidean time in Eq. (\ref{metric2}).

Finally we turn to the task to find quantum states of the scalar
field obeying Eq. (\ref{esch eq}) explicitly. In Ref. \cite{kim2}
the Liouville-Neumann method has been used to construct the
Hilbert spaces for quantum inflatons in the FRW background exactly
for a quadratic potential and approximately for a generic
potential. Similarly we look for the operators that satisfy the
Liouville-Neumann equation
\begin{equation}
i \hbar \frac{\partial}{\partial \tau}
\left\{\matrix{\hat{A}^{\dagger} \cr \hat{A} \cr} \right\}
+ \left[\left\{\matrix{\hat{A}^{\dagger} \cr \hat{A} \cr} \right\}
, \hat{H}_{E} \right] = 0.
\end{equation}
Two independent Liouville-Neumann operators are found
\begin{eqnarray}
\hat{A}^{\dagger} (\tau) &=& - i \Bigl( \varphi_E (\tau)\hat{\pi}_{\phi}
- a^3 (\tau)
\frac{\partial \varphi_E (\tau)}{\partial \tau} \hat{\phi} \Bigr),
\nonumber\\
\hat{A} (\tau) &=&  i \Bigl( \varphi^*_E (\tau) \hat{\pi}_{\phi}
- a^3 (\tau)
\frac{\partial \varphi^*_E (\tau)}{\partial \tau} \hat{\phi} \Bigr),
\end{eqnarray}
where $\varphi_E$ is a complex solution to the equation
\begin{equation}
\frac{\partial^2 \varphi_E (\tau)}{\partial \tau^2} + 3
\Bigl(\frac{\partial a(\tau)/ \partial \tau}{a(\tau)} \Bigr)
\frac{\partial \varphi_E (\tau)}{\partial \tau} - \frac{\delta^2
V(\hat{\phi})}{\delta \hat{\phi}^2 } \varphi_E (\tau) = 0.
\label{aux eq}
\end{equation}
Gaussian states are obtained by taking the expectation value of
Eq. (\ref{aux eq}) with respect to the ground state defined by
$\hat{A} (\tau) \vert 0 (\tau) \rangle = 0$, and by solving the
following equation
\begin{equation}
\frac{\partial^2 \varphi_E (\tau)}{\partial \tau^2}
+ 3 \Bigl(\frac{\partial a(\tau)/\partial \tau}{a(\tau)} \Bigr)
\frac{\partial \varphi_E (\tau)}{\partial \tau}
 - \langle 0 (\tau) \vert \frac{\delta^2 V(\hat{\phi})}{\delta
\hat{\phi}^2 } \vert 0 (\tau) \rangle \varphi_E (\tau) = 0.
\label{36}
\end{equation}
It should be noted that Eq. (\ref{aux eq}) can also be obtained by
the Wick rotation of the Lorentzian equation
\begin{equation}
\frac{\partial^2 \varphi_L (t)}{\partial t^2} + 3
\Bigl(\frac{\partial a(t)/ \partial t}{a(t)} \Bigr) \frac{\partial
\varphi_L (t)}{\partial t} + \frac{\delta^2 V(\hat{\phi})}{\delta
\hat{\phi}^2 } \varphi_L (t) = 0.
\end{equation}
Note also that the inverted potential in Eq. (\ref{36}) can be
obtained through mean-field approximation and Wick rotation of the
Heisenberg equation of motion in the Lorentzian region
\begin{equation}
\frac{\partial^2 \hat{\phi}_L}{\partial t^2}
+ 3 \Bigl(\frac{\partial a(t) / \partial t}{a(t)} \Bigr)
\frac{\partial \hat{\phi}_L (t)}{\partial t} +
 \frac{\delta V(\hat{\phi}_L)}{\delta \hat{\phi}_L }  = 0.
\end{equation}
All these aspects are expected in the Wick rotation of quantum
theory in the Minkowski spacetime.

\section{Transformation between Lorentzian and Euclidean
Quantum Geometries}

In the tunneling universe of Fig. 1, the Lorentzian geometry is
sewn to the Euclidean geometry. There should be a matching
condition or surgery of two geometries. Classically to match
smoothly across the boundary the extrinsic curvature should be
continuous across the boundary. In the FRW universe where the
Lorentzian spacetime is connected to the Euclidean spacetime, the
extrinsic curvature is given by $\pi_a$. We also require that the
geometric quantities $a, \pi_a$ and physical quantities $\phi,
\pi_{\phi}$ be continuous. Sometimes all these are meant the
continuity of wave function of the Wheeler-DeWitt equation across
the boundary just as the wave function of a quantum mechanical
system is continuous across the boundary of tunneling barrier.
Though tempted to continue analytically Eq. (\ref{sem ein}) to
describe quantum theory in the Euclidean region, we have seen that
such a prescription does not provide a good picture for quantum
theory particularly for a gravity-matter system.

In the quantum Lorentzian geometry, the quantum theory of the
scalar field in the Euclidean region I is defined in an ad hoc
manner via the non-unitary Schr\"{o}dinger equation. Besides, the
quantum operators $\hat{\pi}_{\phi}$ and $\hat{\phi}$, and all the
quantization rules are defined in the exactly same manner as in
Lorentzian region. This is not obviously a Wick-rotation. On the
other hand, in the quantum Euclidean geometry the oscillatory
behavior of the Wheeler-DeWitt equation in the same region I
enables one to apply the semiclassical quantum gravity approach to
obtain a well-defined quantum theory of the scalar field. To get a
quantum picture for the scalar field in the Euclidean region there
should be a transformation of the Hilbert space constructed in the
quantum Euclidean geometry into that of the quantum Lorentzian
geometry.

In the de Broglie-Bohm interpretation the canonical
momenta are related to the actions
\begin{equation}
\pi_{a} = \frac{\partial S_{L(II)}}{\partial a},
~\pi_{E, a} = \frac{\partial S_{E}}{\partial a}.
\label{mom-act}
\end{equation}
$\pi_a$ is well defined in the region II $(a \geq a_0)$, since
\begin{equation}
\bigl(\pi_{a} \bigr)^2 = - 3\pi m_P^2 a V_{L} (a) \geq 0,
\end{equation}
whereas $\pi_{E,a}$ is well defined in the region I $(a \leq
a_0)$, since
\begin{equation}
\bigl(\pi_{E,a} \bigr)^2 =  3 \pi m_P^2 a V_{E} (a) \geq 0.
\end{equation}
To find the momentum $\pi_a $ of the Lorentzian geometry in the
Euclidean region I we transform back $\pi_{E, a}$ by the inverse
Wick rotation $\tau = -i s$. Hence, momenta in the Lorentzian and
Euclidean geometries are related by the following transformations
\begin{eqnarray}
\pi_{E, a} &=& i \pi_a,
\nonumber\\
\pi_{E, \phi} &=& i \pi_{\phi}.
\label{wick trans}
\end{eqnarray}
By making use of Eqs. (\ref{mom-act}) and (\ref{wick trans}) we
recover the gravitational field wave function (\ref{grav funI}) of
the Lorentzian geometry from that of the Euclidean geometry:
\begin{equation}
\Psi_E (a) = F(a) \exp \Bigl[\pm \frac{i}{\hbar} S_E (a) \Bigr]
\Rightarrow \Psi_L (a) = F(a) \exp \Bigl[\mp \frac{1}{\hbar}
S_{L(I)} (a) \Bigr].
\end{equation}

We turn to the transformation of quantum states of the scalar
field. In the region II the scalar field has the energy
expectation value with respect to the symmetric Gaussian state
$\langle \hat{\phi} \rangle = \phi_c = 0 $
\begin{equation}
\langle \hat{H}_L \rangle = \pi^2 \hbar^2 a^3
\frac{\partial \varphi_{L(II)}^*}{\partial t}
\frac{\partial \varphi_{L(II)}}{\partial t}
+ 2 \pi^2 a^3 \Biggl[\exp \Bigl(\frac{\hbar^2}{2} \varphi_{L(II)}^*
\varphi_{L(II)} \frac{\partial^2 }{\partial \phi_c^2} \Bigr)
-1 \Biggr] V(\phi_c = 0).
\end{equation}
Similarly, in the region I of the Euclidean geometry the energy
expectation value is given by
\begin{equation}
\langle \hat{H}_E \rangle = - \pi^2 \hbar^2 a^3 \frac{\partial
\varphi_{E}^*}{\partial \tau} \frac{\partial \varphi_{E}}{\partial
\tau} + 2 \pi^2 a^3 \Biggl[\exp \Bigl(\frac{\hbar^2}{2}
\varphi_{E}^* \varphi_{E} \frac{\partial^2 }{\partial \phi_c^2}
\Bigr) -1 \Biggr] V(\phi_c = 0). \label{euc exp}
\end{equation}
Thus $\langle \hat{H}_E \rangle$ is the true Wick rotation of
$\langle \hat{H}_L \rangle$. This justifies the fact that the
region I of the Lorentzian geometry coincides with the region
where $V_E$ is positive and the wave function oscillates. The Wick
rotation transforms $H_E$ back into $H_L$ and recovers the
positive signature of the kinetic term. Likewise, Eq. (\ref{sch
eqI}) is the Wick rotation of Eq. (\ref{esch eq}). Therefore, in
the region I all quantum states of the scalar field in the
Lorentzian geometry are obtained by Wick rotating those in the
Euclidean geometry.

How can we find directly quantum states in the Lorentzian
geometry? For this purpose we should find the quantization rule in
the region I
\begin{equation}
\bigl[\hat{\phi}, \hat{\pi}_{\phi} \bigr] = \hbar,
\end{equation}
which follows from Eq. (\ref{mom-act}) and
the standard quantization in the Euclidean geometry
\begin{equation}
\bigl[\hat{\phi}_E, \hat{\pi}_{E, \phi} \bigr] = i \hbar.
\end{equation}
Though not firmly established, we may use the non-unitary version
of the Liouville-Neumann equation
\begin{equation}
\hbar \frac{\partial}{\partial s} \hat{\cal O}
+ \bigl[\hat{\cal O}, \hat{H}_L \bigr] = 0.
\end{equation}
Two operators are found
\begin{eqnarray}
\hat{A}_+ (s) &=& \varphi_{L(I)} (s) \hat{\pi}_{\phi}
- a^3 \frac{\partial \varphi_{L(I)}(s)}{\partial s}
\hat{\phi},
\nonumber\\
\hat{A}_- (s) &=& - \varphi_{L(I)}^* (s) \hat{\pi}_{\phi}
+ a^3 \frac{\partial \varphi_{L(I)}^*(s)}{\partial s}
\hat{\phi},
\end{eqnarray}
where $\varphi_{L(I)}$ is a complex solution to the equation
\begin{equation}
\frac{\partial^2 \varphi_{L(I)}(s)}{\partial s^2} + 3 \Bigl(
\frac{\partial a/ \partial s}{a}\Bigr) \frac{\partial
\varphi_{L(I)}(s)}{\partial s} + \frac{\delta^2 V(\hat{\phi})
}{\delta \hat{\phi}^2} \varphi_{L(I)}(s) = 0. \label{aux eq2}
\end{equation}
The operators $\hat{A}_+ (s)$ and $\hat{A}_- (s)$ play the same
role as $\hat{A}^{\dagger} (\tau)$ and $\hat{A} (\tau)$,
respectively. Note that Eq. (\ref{aux eq2}) is the Wick rotation
of Eq. (\ref{aux eq}).

Finally we discuss the interpretation of a bounce solution of the
semiclassical Einstein equation (\ref{sem ein2-2}) in the
Euclidean geometry in terms of temperature for a tunneling regime
in the Lorentzian geometry \cite{brout}. Without the back-reaction
of matter, Eq. (\ref{sem ein2-2}) has a periodic solution
\begin{equation}
a (\tau) = \sqrt{\frac{3}{\Lambda}} \cos
\Bigl(\sqrt{\frac{\Lambda}{3}} \tau \Bigr). \label{per sol}
\end{equation}
The periodic solution (\ref{per sol}) is the analytic continuation
of the de Sitter solution to the semiclassical Einstein equation
(\ref{sem ein1-1}) in the Lorentzian geometry
\begin{equation}
a (t) = \sqrt{\frac{3}{\Lambda}} \cosh
\Bigl(\sqrt{\frac{\Lambda}{3}} t \Bigr). \label{de sitter}
\end{equation}
When we adopt the standard interpretation of finite temperature
fields in the Minkowski spacetime, the period of Eq. (\ref{per
sol}) corresponds to an inverse temperature
\begin{equation}
\tau = \frac{1}{T} = \frac{2 \pi}{\sqrt{\frac{\Lambda}{3}}}.
\end{equation}
This coincides with the temperature for the de Sitter spacetime
from other methods. However, with the back-reaction (\ref{euc
exp}), Eq. (\ref{sem ein2-2}) reads that
\begin{equation}
\Biggl(\frac{\frac{\partial a}{\partial \tau}}{a} \Biggr)^2 -
\frac{1}{a^2} + \frac{\Lambda}{3} =  \frac{4 \pi}{3 m_P^2}
\Biggl\{\hbar^2 \frac{\partial \varphi_{E}^*}{\partial \tau}
\frac{\partial \varphi_{E}}{\partial \tau} + 2 \Biggl[\exp
\Bigl(\frac{\hbar^2}{2} \varphi_{E}^* \varphi_{E} \frac{\partial^2
}{\partial \phi_c^2} \Bigr) -1 \Biggr] V(\phi_c = 0) \Biggr\}.
\label{sem ein2-3}
\end{equation}
The task to find the temperature for the gravity-matter system is
equivalent to solving both Eqs. (\ref{sem ein2-3}) and (\ref{aux
eq}) or (\ref{36}) and finding a periodic solution. This requires
a further study \cite{kim3}.

\section{Conclusion}

We have studied quantum field theory of matter in the universe
undergoing a topology change from the Euclidean region into the
Lorentzian region. It is shown that the semiclassical gravity
derived from canonical quantum gravity provides a consistent
scheme for quantum field theory in such topology changing
universes. The Lorentzian and Euclidean regions of spacetime are
classified according to the behavior of the wave function of the
gravitational field with the quantum back-reaction of matter
included. In the Lorentzian region the gravitational wave function
oscillates. Provided a cosmological time is properly chosen along
the trajectory of oscillating gravitational wave function, the
semiclassical Einstein equation has the same form as the classical
Einstein equation with the quantum back-reaction of matter as a
source. The time-dependent Schr\"{o}dinger equation is identical
to canonical quantum field equation.

On the other hand, in the Euclidean region the gravitational wave
function shows either an exponentially growing or decaying
behavior or a superposition of them. One may derive the
semiclassical Einstein equation and time-dependent Schr\"{o}dinger
equation in the sense of analytic continuation. However, it is
found that the Schr\"{o}dinger equation evolves like a diffusion
equation, not preserving unitarity, and the quantization rule
differs from the usual one in the Lorentzian spacetime.

In order to construct a consistent quantum theory of matter field
we have proposed a scheme in which the gravity-matter system is
Wick-rotated in the Euclidean region of spacetime and the
semiclassical (quantum) gravity is derived from the Wheeler-DeWitt
equation for the Wick-rotated Euclidean geometry. The
time-dependent Schr\"{o}dinger equation is well defined as in the
Lorentzian region and quantum states are found using the
Liouville-Neumann method. Finally these quantum states are
transformed via the inverse Wick rotation back into those of the
Lorentzian geometry.

This quantum field theory applies to the Universe that emerged
quantum mechanically from a Euclidean region of spacetime. It
would be interesting to see the physical consequences of the
quantum fields for different boundary conditions of the Universe.
Another physically interesting problem requiring a further study
is to find the period solution of both the semiclassical Einstein
equation and matter field equation in the Euclidean geometry and
to interpret the period as the inverse temperature for the
gravity-matter system in the tunneling regime of the Lorentzian
geometry \cite{brout}.

\acknowledgements

The author wishes to acknowledge the financial support of
the Korea Research Foundation under contract No. 1998-001-D00364
and through BSRI Program under contract No. 1998-015-D00129.

\begin{figure}
\begin{center}
\epsfxsize=4.0in \epsffile{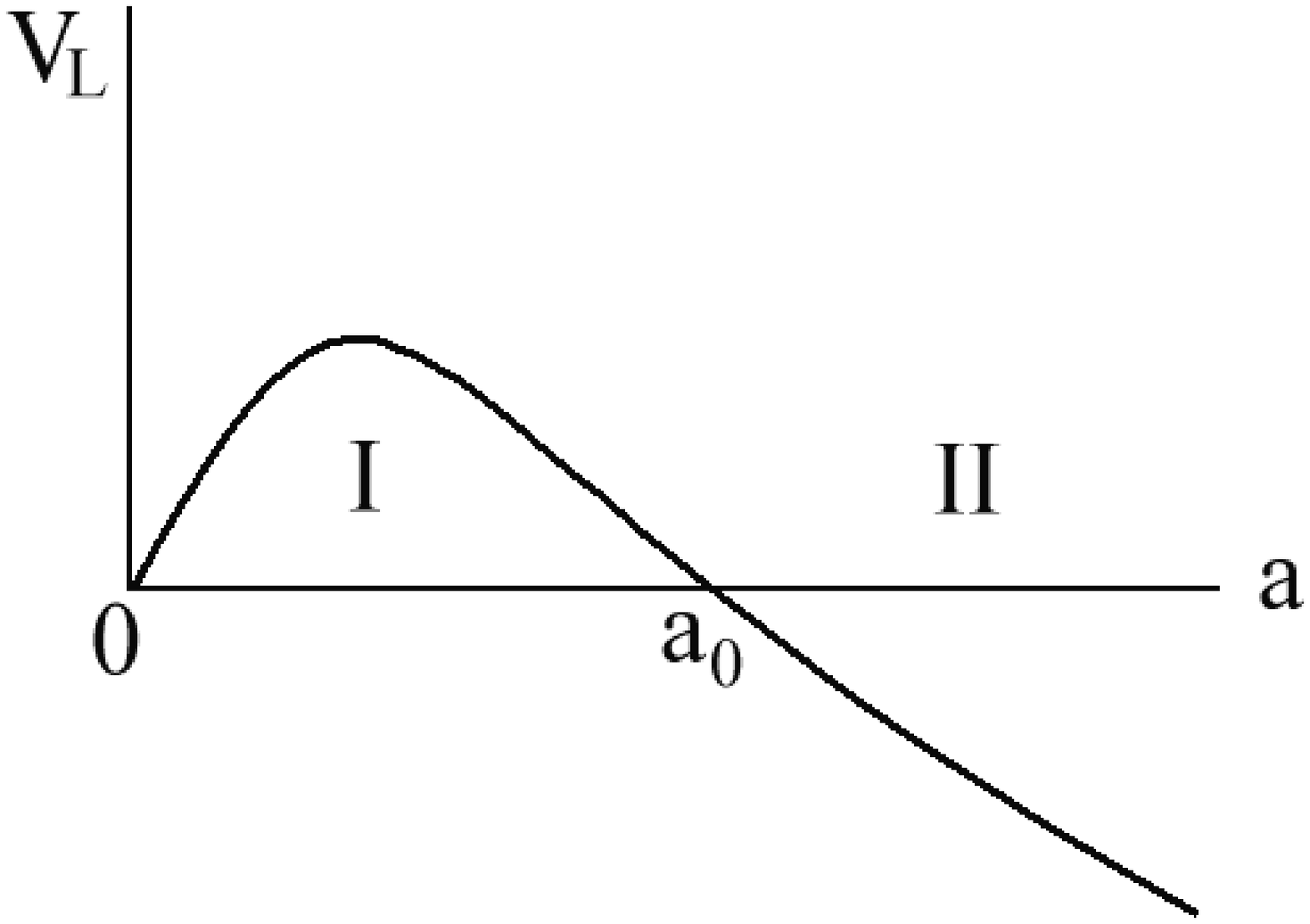} \vspace{5cm} \caption{The
effective potential $V_L$ vs. $a$ in the Lorentzian geometry. The
gravitational field exhibits an exponential behavior in the
Euclidean region I but oscillates in the Lorentzian region II.}
\end{center}
\end{figure}

\end{document}